\documentclass[prl,twocolumn,preprintnumbers,amsmath,amssymb,groupeaddress,superscriptaddress]{revtex4}
\usepackage[pdftex]{graphicx}% Include figure files
\usepackage{dcolumn}% Align table columns on decimal point
\usepackage{bm}% bold math
\usepackage[latin1]{inputenc}

\newcommand{\ud}[1]{{#1^{\dagger}}}

\newcommand{\mean}[1]{\langle#1\rangle}

\begin{document}

%\preprint{APS/123-QED}

\title{Lasing in Strong Coupling}

\author{F. P. Laussy} 
\affiliation{Walter Schottky Institut, Technische Universit\"at M\"unchen, Am Coulombwall 3, 85748 Garching, Germany}
%\email{fabrice.laussy@gmail.com}
\author{E. del Valle} 
\affiliation{Physikdepartment, TU M\"unchen, James-Franck-Str. 1, 85748 Garching, Germany}
%\email{elena.delvalle.reboul@gmail.com}
\author{J. J. Finley} 
\affiliation{Physikdepartment, TU M\"unchen, James-Franck-Str. 1, 85748 Garching, Germany}
\affiliation{Walter Schottky Institut, Technische Universit\"at M\"unchen, Am Coulombwall 3, 85748 Garching, Germany}

\date{\today}% It is always \today, today,

\begin{abstract}
  An almost ideal thresholdless laser can be realized in the
  strong-coupling regime of light-matter interaction, with Poissonian
  fluctuations of the field at all pumping powers and all intensities
  of the field. This ideal scenario is thwarted by quantum
  nonlinearities when crossing from the linear to the stimulated
  emission regime, resulting in a universal jump in the second order
  coherence, which measurement could however be used to establish a
  standard of lasing in strong coupling.
\end{abstract}

%\pacs{}
\maketitle

``Lasing'' as a general concept is any process that generates a
coherent field (we will assume light).  A single emitter can be used
for that purpose, provided that it is in the \emph{strong coupling}
with the field~\cite{kavokin_book11a}, in which case the interaction
is reversible and can pile-up coherently a very large number of
photons through Rabi oscillations~\cite{mu92a}. In contrast,
conventional lasers operate in the \emph{weak-coupling regime}, where
the interaction is perturbative. This demands a large number~$N\gg1$
of emitters to generate a sizable output. The inversion of population
of this ensemble leads to a pumping threshold. With a single emitter,
if the spontaneous emission rate into other modes than the cavity is
small, the growth in the population of photons appears to exhibit no
threshold~\cite{rice94a}.  Even though a proper coherent state is not
formed due to the uncertainty in the phase (blurred by the interaction
with the environment~\cite{karlovich01a}), the distribution converges
to the same Poissonian statistics.  Most lasers find their
applications in their high intensity and/or narrow (so-called
``pencil'') beam, but from a fundamental point of view,
\emph{coherence} as defined by Glauber~\cite{glauber63c}, i.e., as
autocorrelation functions~$N_a[n]=\langle\ud{a}^na^n\rangle$ of the
field that factor out, is what endows lasing with its cleanest
definition.

The strong-coupling regime is not interesting only for its lasing
properties. It presents particular quantum nonlinearities that arise
in the fully quantized theory~\cite{jaynes63a,cirac91a}.
Experimentally, entering the strong coupling regime at the single
excitation level is technically demanding, both in the case of atoms
and of \emph{artificial atoms} (superconducting qubits and quantum
dots). This requires a well isolated system with a high oscillator
strength for the emitter and quality factor for the cavity, matching
the emitter and the chosen cavity mode both spatially and
spectrally. It was only in 2004 that all these requirements were met
and strong coupling was achieved for a variety of systems: a single
and the same trapped atom in an optical cavity~\cite{boca04a}, a
superconducting qubit in a superconducting transmission line
resonator~\cite{wallraff04a} and a quantum dot in a semiconductor
microcavity~\cite{reithmaier04a,yoshie04a,peter05a}. Signatures of
strong coupling at the two excitation level have also been reported
subsequently in all these systems~\cite{boca04a,fink08a,kasprzak10a},
as well as one-emitter
lasing~\cite{mckeever03a,astafiev07a,nomura10a}, showing that they can
be considered as two-level systems.

Theoretically, much work has addressed the steady state properties of
the one-emitter laser (field intensity, statistics, population
inversion\dots) through the different regions of pumping (quantum,
lasing and
quenching)~\cite{mu92a,ginzel93a,loffler97a,jones99a,benson99a,karlovich01a,kilin02a,clemens04a,karlovich08a,delvalle09a,poddubny10a,andre10a,ritter10a,auffeves10a,delvalle10d,lsc_gartner11a,arXiv_delvalle11a}.
In this text, we focus on the transition from strong coupling at low
excitations to lasing sustained by the single-emitter. We show how,
when conditions are optimum for strong-coupling, all observables
(intensity and coherence) tend toward a cancellation of all transients
and thresholds. We find, however, a universal ``jump'' when bridging
between these two limits, that forbids the realization of an ideal
thresholdless laser. We discuss how this can be taken advantage of.

Lasing in strong coupling is described at its most fundamental level
by the coupling between a two-level system~$\sigma$ and a cavity
mode~$a$ in a dissipative environment that leads to a master
equation
$\partial_t\rho=i[H_\mathrm{JC},\rho]+\{\frac{\gamma_a}{2}\mathcal{L}_a+\frac{\gamma_\sigma}{2}\mathcal{L}_\sigma+\frac{P_\sigma}{2}\mathcal{L}_\ud{\sigma}\}\rho$,
%+\frac{\gamma_\phi}{2}\mathcal{L}_{\ud{\sigma}\sigma}
where $\mathcal{L}_c\rho=(2c\rho\ud{c}-\ud{c}c\rho-\rho\ud{c}c)~$ is
the Lindblad term associated with decay ($\gamma_a$, $\gamma_\sigma$),
and pumping ($P_\sigma$), and $H_\mathrm{JC}$ is the celebrated
Jaynes--Cummings Hamiltonian, at the heart of the quantum dynamics:
$H_\mathrm{JC}=g(\sigma^\dagger a+a^\dagger \sigma)$.
%
%\begin{equation}
%  \label{eq:WedApril27221736CEST2011}
%  H_\mathrm{JC}=\omega_a a^\dagger a+\omega_\sigma\sigma^\dagger\sigma+g(\sigma^\dagger a+a^\dagger \sigma)\,.
%\end{equation}
%
The steady state can be expressed completely in terms of the photon
correlators, which obey the equations~\cite{delvalle09a}:
\begin{multline}
  \label{eq:MoMar22173241WET2010}
  \Big[1+\frac{\Gamma_\sigma+(2n-1)\gamma_a}{\kappa_\sigma}+\frac{n\gamma_a}{\Gamma_\sigma+(n-1)\gamma_a}-\frac{2P_\sigma}{\Gamma_\sigma+n\gamma_a}\Big]N_{a}[n]\\
  =\frac{nP_\sigma}{\Gamma_\sigma+(n-1)\gamma_a}N_a[n-1]-\frac{2\gamma_a}{\Gamma_\sigma+n\gamma_a}N_a[n+1]\,.
\end{multline}
where we have introduced
$\Gamma_\sigma=\gamma_\sigma+P_\sigma$ % , respectively, the effective cooperativity
% and coupling:
% %
% \begin{subequations}
%   \label{eq:MonMar28154711CEST2011}
%   \begin{align}
%     C_\mathrm{eff}[m]=&\frac{4 (g_\mathrm{eff}[m])^2}{\gamma_a(\Gamma_\sigma+(2m-1)\gamma_a)} \,,\\
%     g_\mathrm{eff}[m]=&\frac{g}{\sqrt{1+\big[2\Delta/(\Gamma_\sigma+(2m-1)\gamma_a)\big]^2}}
%     \,.
%   \end{align}
% \end{subequations}
% %
% We also introduced for convenience
and the Purcell rate of transfer of population from emitter to the
cavity mode:
\begin{equation}
  \label{eq:MonMar28155703CEST2011}
  \kappa_\sigma={4g^2}/{\gamma_a}\,.
\end{equation}

The main observables of interest are the cavity population,
$n_a=N_a[1]$, directly linked to the intensity emitted by the device
through $I=\gamma_an_a$, and the $n$th-order coherence function
$g^{(n)}(0)=N_a[n]/n_a^n$, especially the second order one, $g^{(2)}$,
measured by photon-counting coincidences at zero time delay. The
probability of the emitter to be in the excited state
$n_\sigma=\mean{\ud{\sigma}\sigma}$ is a dependent variable
($n_\sigma=(P_\sigma-\gamma_an_a)/\Gamma_\sigma$), which we thus do
not need to consider any further. These equations can be solved to
very good approximation~\cite{arXiv_delvalle11a}. Two cases, $i=1, 2$,
when the field intensity scales linearly with pumping at a rate of
growth~$C_i$, are of interest:
\begin{equation}
  \label{eq:FriMay20201139CEST2011}
  n_a=C_iP_\sigma\,.
\end{equation}

\begin{figure}[t]
  \centering
  \includegraphics[width=.8\linewidth]{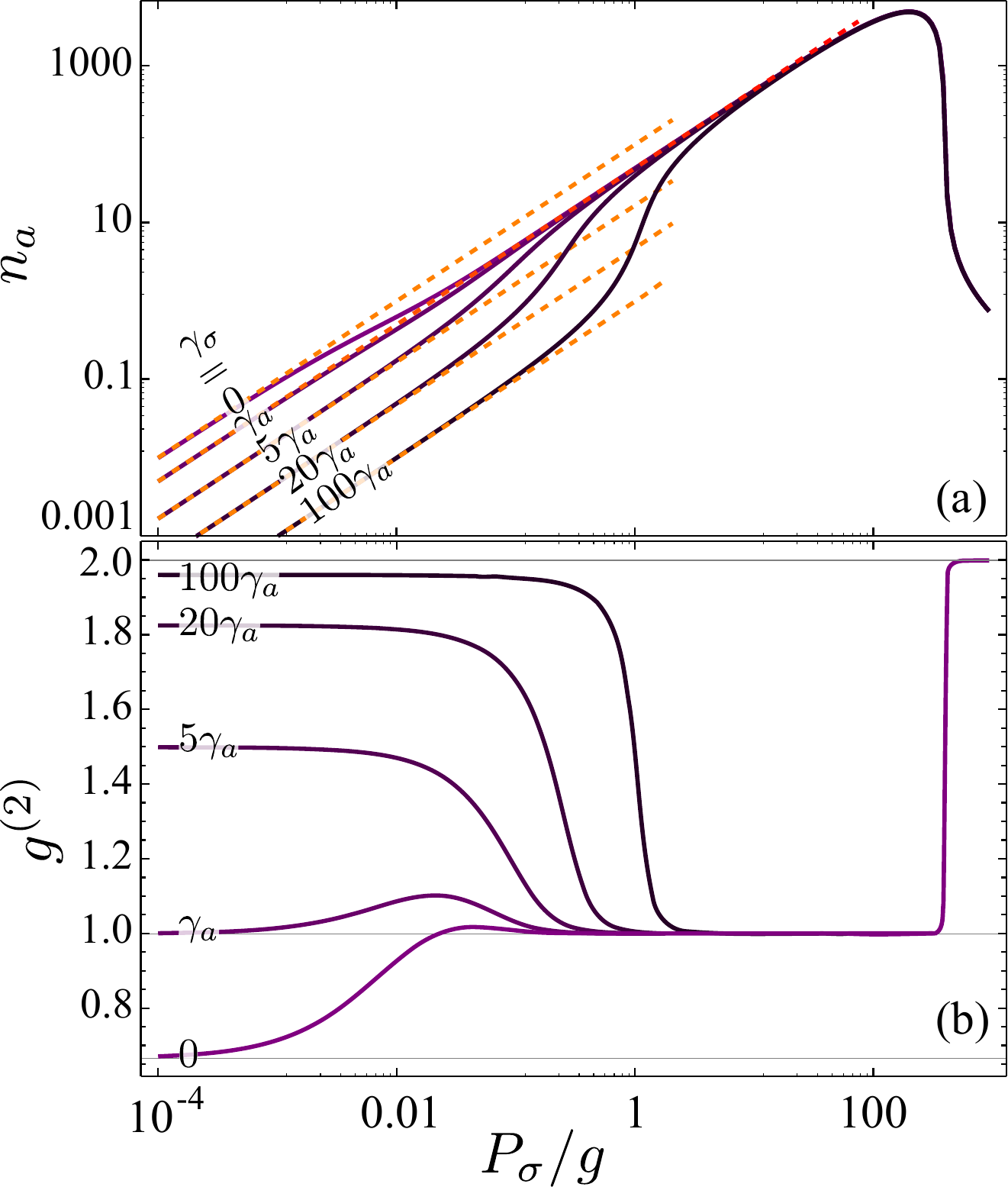}
  \caption{(Colour online) (a) Cavity population $n_a$ and (b) second
    order correlation $g^{(2)}$ for various $\gamma_\sigma$ (with
    $\gamma_a/g=10^{-2}$).  The linear variation $n_a=C_1P_\sigma$ is
    superimposed in dashed orange lines. In the stimulated emission
    regime, all lines converge to the same one (red). The second
    uppermost curve where $C_1=C_2$, that covers both regimes, is the
    closest approximation to an ideal thresholdless laser in
    strong-coupling. It however deviates slightly in the intermediate
    region.  This deviation becomes compelling in $g^{(2)}$, where it
    arises as a bunching of photons when turning a perfect Poissonian
    distribution at small pump into another one at high pump. This
    curve, which is universal, is magnified in
    Fig.~\ref{fig:TueMay31115403CEST2011}.}
  \label{fig:WedMay18011335CEST2011}
\end{figure}

In the ``linear'' regime---where only the first rung of the
Jaynes--Cummings ladder is occupied---we find:
\begin{equation}
  \label{eq:ThuMay12181806CEST2011}
    C_1\approx\frac{\kappa_\sigma}{\kappa_\sigma+\gamma_\sigma}\frac{1}{\gamma_a+\gamma_\sigma}P_\sigma \,.
\end{equation}
This is shown in Fig.~\ref{fig:WedMay18011335CEST2011}(a) at low pump,
where $n_a$ follows the dashed straight lines, given by
Eq.~(\ref{eq:ThuMay12181806CEST2011}).  In the lasing region, the
field intensity also scales linearly with pumping, but this time with
a rate independent of $\gamma_\sigma$, since spontaneous processes are
completely dominated by stimulated ones:
\begin{equation}
  \label{eq:FriMay20200841CEST2011}
%  C_2\approx\frac{1}{2\gamma_a}\,.
  C_2\approx{1}/(2\gamma_a)\,.
\end{equation}
This is the region in Fig.~\ref{fig:WedMay18011335CEST2011}(a) where
all lines converge (since $\gamma_a$ is constant). There is therefore
a ``jump'' $\mathcal{J}$ between the two rates of efficiency in the
transition from the linear to the lasing regime:
\begin{equation}
  \label{eq:FriMay20200916CEST2011}
%  \mathcal{J}=\ln\left(\frac{C_2}{C_1}\right)\approx\ln\frac{\gamma_a+\gamma_\sigma}{2\gamma_a}\,.
  \mathcal{J}=\ln\left({C_2}/{C_1}\right)\approx\ln(\gamma_a+\gamma_\sigma)-\ln(2\gamma_a)\,,
\end{equation}
which becomes exact when $\kappa_\sigma\gg\gamma_\sigma$.  This jump
changes sign when $\gamma_\sigma=\gamma_a$, which is the condition
that maximises the strong-coupling criterion:
\begin{equation}
  \label{eq:TueMay31172337CEST2011}
  4g>|\gamma_a-\gamma_\sigma|\,.
\end{equation}
Cases that satisfy Eq.~(\ref{eq:TueMay31172337CEST2011}) with
$\gamma_\sigma<\gamma_a$ result in a \emph{drop down} of the
efficiency of pumping when crossing from the linear to the quantum
regime, while cases $\gamma_\sigma>\gamma_a$ undergo a \emph{bounce
  up}, as stimulated emission overcomes spontaneous emission according
to the conventional lasing scenario. The drop down is maybe more
surprising. It is maximum when $\gamma_\sigma=0$ (no spontaneous
emission) in which case $\mathcal{J}=\ln(1/2)$, the factor~$1/2$ being
linked to the inversion of population (in the lasing region,
$n_\sigma=1/2$). When the inequality~(\ref{eq:TueMay31172337CEST2011})
is maximum, the situation of an ideal thresholdless laser would seem
to be realized, namely, the light field intensity increases linearly
with pumping throughout the entire excitation scheme (until
quenching). This is not entirely true, however, since in between the
linear regime and the lasing regime lies what we will call the
``\emph{quantum regime}'', where the dynamics involving the first few
rungs of the Jaynes--Cummings ladder disrupt anyway the zero-jump
between the two linear relationships when $C_1=C_2$. This is shown in
Fig.~\ref{fig:WedMay18011335CEST2011}(a) where one sees that the case
$\gamma_\sigma=\gamma_a$ accounts for both the linear and the lasing
regions with the same line, with a small deviation in the intermediate
region.

This is even more apparent when considering the statistics~$g^{(2)}$,
in Fig.~\ref{fig:WedMay18011335CEST2011}(b). As photon correlators
follow $N_a[n]\propto P_\sigma^n$ at vanishing pump, a finite value
for all $g^{(n)}$ is assured independently of the truncation used to
solve Eq.~(\ref{eq:MoMar22173241WET2010}). We thus obtain the exact
expression for the general coherence function in the limit of
vanishing pump:
%
%\begin{widetext}
\begin{multline}
    \label{eq:FriMay13201821CEST2011}
    % g_0^{(n)}=\lim_{P_\sigma\rightarrow
    %   0}\frac{N_a[n]}{n_a^n}=ng_0^{(n-1)}
    % \frac{\kappa_\sigma (\gamma_a +\gamma_\sigma)+\gamma_\sigma
    %   (\gamma_a+\gamma_\sigma)}{\kappa_\sigma \big[
    %   (2n-1) \gamma_a +\gamma_\sigma\big]+
    %   \big[(n-1)\gamma_a+\gamma_\sigma\big]
    %   \big[(2n-1)\gamma_a+\gamma_\sigma\big]} \,,
    g_0^{(n)}=\lim_{P_\sigma\rightarrow 0}\frac{N_a[n]}{n_a^n}=\\ng_0^{(n-1)}
    \frac{\kappa_\sigma+\gamma_\sigma}{\kappa_\sigma+\gamma_\sigma+(n-1)\gamma_a}
    \frac{\gamma_a+\gamma_\sigma}{(2n-1)\gamma_a+\gamma_\sigma}\,,
\end{multline}
%\end{widetext}
%
Starting from
$g_0^{(1)}=1$.  % The maximum value this can reach is $n!$, in which
% case the photons are completely bunched with a thermal distribution.
% This is the case if the fraction in
% Eq.~(\ref{eq:FriMay13201821CEST2011}) is 1. Otherwise, this fraction
% containes all the quantum and coherent properties of the system.
The minimum value $g_0^{(n)}$ can take is~$0$, the case of perfect
antibunching. % For large $n$, light always becomes antibunched,
% $g_0^{(n)}/g_0^{(n-1)}\sim 1/(2(1+n))\rightarrow 0$, as one would
% expect from the incoherent excitation.
In the very strong coupling regime (where $\kappa_\sigma$ is the
largest parameter), $g_0^{(2)}$ can be approximated as:
\begin{equation}
  \label{eq:MonMay23015052CEST2011}
  g_0^{(2)}\approx2(\gamma_a +\gamma_\sigma)/(3 \gamma_a +\gamma_\sigma)\,,
\end{equation}
which is always between 2/3 and~2, as shown in
Fig.~\ref{fig:WedMay18011335CEST2011}(b).  This result has also been
recently obtained by a continuous fraction
expansion~\cite{lsc_gartner11a}.  The condition for $g_0^{(2)}=1$,
which separates the bunching ($>1$) from the antibunching ($<1$)
behaviour, is again $\gamma_\sigma=\gamma_a$, the same criterion as
the one that aligns the two linear growths. All higher order
correlators, Eq.~(\ref{eq:FriMay13201821CEST2011}), satisfy
$g_0^{(n)}=1$ in this case, showing that the state is exactly
Poissonian or, in the sense of Glauber, perfectly coherent.

In the second order statistics, the passage through the quantum region
is however markedly located as a ``bump'' in an otherwise constant
$g^{(2)}=1$. There is some interest in having a stable light source
with a pinned fluctuation of its statistics as Poissonian for all
intensities, even those much below unity.  The ideal thresholdless
laser would be such that for all pumping powers (below quenching), its
coherence would be that of a laser. Two different mechanisms account
for this Poissonian statistics, though: at low pumping, by maximising
strong coupling; at large pumping, by stimulated emission overtaking
spontaneous emission.

The threshold of a conventional laser is quantified by its $\beta$
factor, which is the closer to one the lower the threshold, a concept
that has been extended to the one-atom laser~\cite{auffeves10a}:
$\beta=[{\kappa_\sigma}/(\kappa_\sigma+\gamma_\sigma)][{\gamma_a}/(\gamma_a+\gamma_\sigma)]$.
%
% \begin{equation}
%   \label{eq:FriMay20210015CEST2011}
%   \beta=\frac{\kappa_\sigma}{\kappa_\sigma+\gamma_\sigma}\frac{\gamma_a}{\gamma_a+\gamma_\sigma}\,.
% \end{equation}
%
In our approximation of $\kappa_\sigma\gg\gamma_\sigma$, $\beta$ is
related to our jump between the linear increases of the single-photon
and stimulated emission lasing regimes as
$\mathcal{J}=\ln(1/(2\beta))$. The $\beta$ factor is the fraction of
emission in the lasing mode (the cavity), which is stimulated, over
other channels of emission, most importantly spontaneous emission
which is always present, at least in weak coupling.  Strong coupling
being this regime where spontaneous emission becomes a reversible
process, we argue that the definition $\beta=1$, or
$\mathcal{J}=-\ln(2)$, suits best weak-coupling lasers and that in
strong-coupling, $\beta=1/2$ or $\mathcal{J}=0$ is the closest, albeit
non-ideal, approximation to thresholdess lasing operation. It is also
conceptually appealing that lasing in strong coupling is best realized
when strong coupling itself is optimum,
Eq.~(\ref{eq:TueMay31172337CEST2011}).  The wider picture covering
both the quantum and classical regimes also reveals different types of
thresholds, namely, from quantum ($g^{(2)}<1$) to classical
($g^{(2)}=1$) statistics when $\gamma_\sigma<\gamma_a$, and from
thermal noise ($g^{(2)}>1$) to classical statistics, which is the
conventional case, when $\gamma_\sigma>\gamma_a$. The intermediate
situation where $\gamma_a=\gamma_\sigma$ bridges between Poissonian
statistics on both sides. If one would assume the efficiency of growth
of the intensity as the criterion for lasing, the negative-jump would
yield an ``anti-threshold'' where stimulated emission spoils the
efficiency of cavity population, strong-coupling being more efficient.
This jump neatly and fundamentally separates two regions that differ
only by the fact that $n_a<1$ in the former case and $n_a>1$ in the
latter, but are otherwise sharing the same growth of the photon
intensity with pumping and Poissonian statistics, that is, both
displaying the two main features of a laser. It is therefore adequate
to denominate them both as lasing. We propose the denominations of
``single-photon lasing'' and ``stimulated-emission lasing'' to term
the two sides of the quantum regime.  The terminology of a
``single-photon laser'', seemingly contradictory in terms,
nevertheless restores the concept of coherence as the chief
characteristic of lasing, since this is precisely \emph{not} the large
intensity (thus, the large number of photons) that characterize
lasing, but the fact that the very scarce photons emitted are
uncorrelated the one from the others, in stark contrast with a natural
source where independent events leads to bunched
photons~\cite{hanburybrown56a}. The same applies to the terminology of
``stimulated-emission lasing'' which is not a pleonasm in a modern
understanding of lasing, where the mechanism is disconnected from its
product.

% This terminology being redundant with the LASER
% acronym, we shall refer to the ``linear regime'' and ``lasing regime''
% instead, and term ``quantum'' regime the intermediate region
% separating them.

\begin{figure}[t] 
  \centering 
\includegraphics[width=\linewidth]{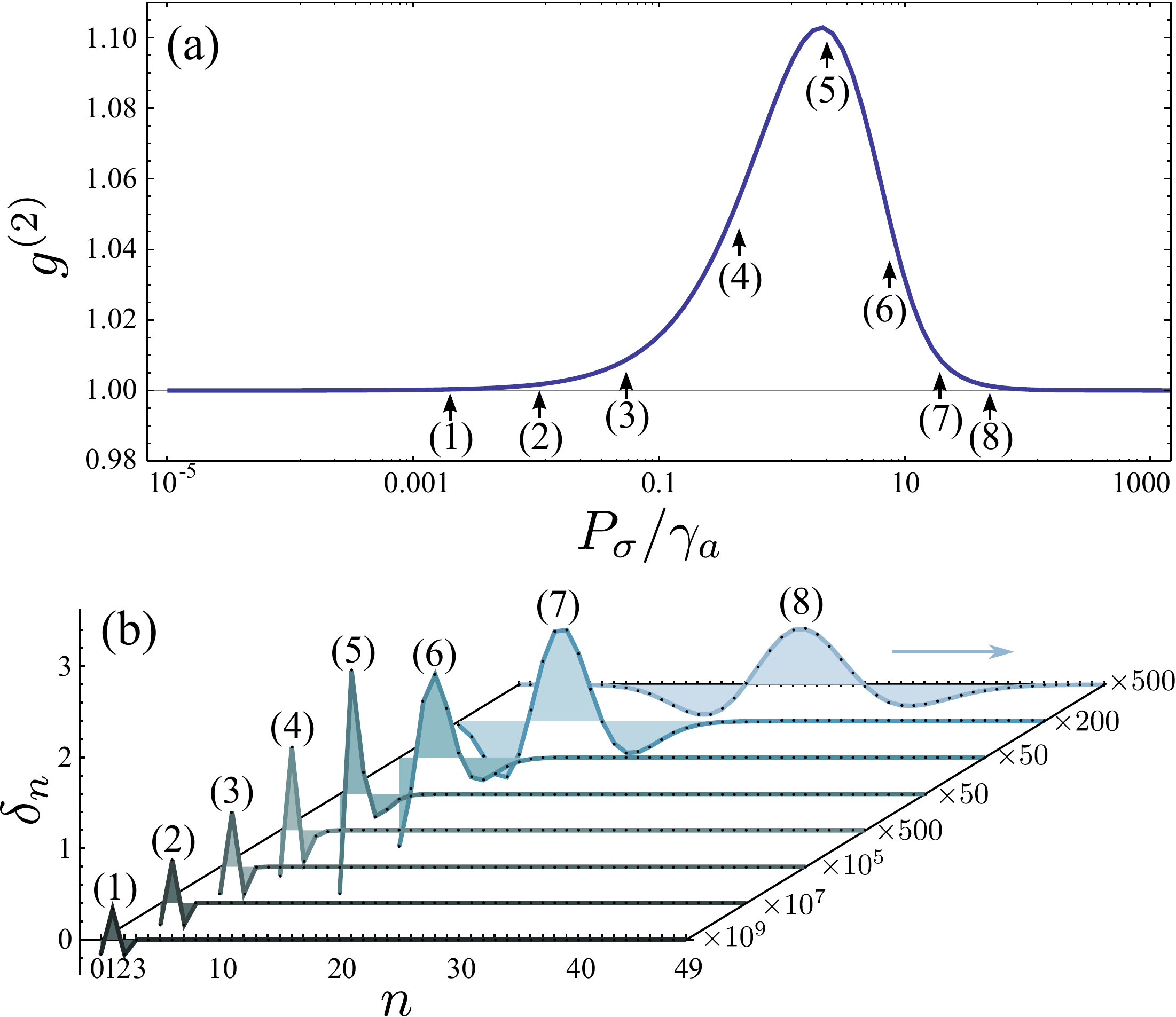}
\caption{(a) universal curve for $g^{(2)}$ when going from one-photon
  lasing to stimulated emission lasing and, (b), deviation of the
  statistics realized from a Poissonian distribution,
  Eq.~(\ref{eq:TueMay31234340CEST2011}), for the points marked by
  arrows in~(a). The maximum value $\approx1.10282$ is the same for
  any system realizing lasing in strong-coupling. $\delta_n$ is
  magnified by the value shown on the right hand side.}
\label{fig:TueMay31115403CEST2011}
\end{figure} 

The most remarkable feature of the transition between these two types
of lasing is that it is universal. This follows from the strong
coupling limit, where the term featuring $\kappa_\sigma^{-1}$ in
Eq.~(\ref{eq:MoMar22173241WET2010}) becomes negligible, in which case
the shape is invariant for the dimensionless parameters
$P=P_\sigma/\gamma_a$ and $\gamma=\gamma_\sigma/\gamma_a$, for all
values of~$g$. It is shown in Fig.~\ref{fig:TueMay31115403CEST2011},
along with the physical origin of this fluctuation in statistics,
displayed as the difference between the distribution~$p(n)=\langle
n|\rho|n\rangle$ realized in the system and the ideal Poissonian
statistics, with $\bar n=\sum_nnp(n)$:
\begin{equation}
  \label{eq:TueMay31234340CEST2011}
  \delta_n=p(n)-e^{-\bar n}\bar n^n/n!\,.
\end{equation}
In the one-photon lasing region (1--3) in
Fig.~\ref{fig:TueMay31115403CEST2011}, the system is forced in the
lowest rung $n=1$, resulting in lower probabilities to have two
photons than in an ideal laser of the corresponding intensity~$\bar
n$.  This imbalance grows linearly and, in the transition region
(4--7), it spreads over many rungs, with excess of photons nearby the
maximum of the distribution while neighbouring rungs are depleted to
compensate. In the stimulated emission lasing region (8), this
perturbation in statistics propagates along the ladder at the same
time as it vanishes, recovering exact Poissonian fluctuations at high
intensities.

The curve becomes not-universal anymore but specifics to the system
parameters when strong coupling is not good enough. The shape then
deviates from that plotted and reaches different (lower) values of its
maximum. Interestingly, this occurs when the lasing regime established
by stimulated emission (after the bump) is no longer reached, that is,
no plateau is fully formed where Poissonian statistics is maintained
over a range of pumping. We place it at roughly
$\gamma_a\approx0.1g$. This shows that the transition is really a
fundamental bridge between the two types of lasing, that disappears if
and only if this crossover is not fully realized.

For good enough strong-coupling, universality implies in particular
that all systems exhibit the same maximum in
$g^{(2)}$. Numerically, we estimate these lowest possible values by
which the system surpass Poissonian statistics to
be% ~\footnote{There might be exact expressions for these
% constants, given that an exact series can be given for $g^{2}$ as a
% function of~$P$, but given its slow rate of convergence and its
% complexity we could not obtain it~\cite{lsc_delvalle11a}.}
:
\begin{align*}
  g^{(2)}\approx1.01816\,,\quad\text{at $P_\sigma\approx4.5989\gamma_a$ when $\gamma_\sigma=0$}\,,\\%    \label{eq:SunMay22230900CEST2011}\\
  g^{(2)}\approx1.10282\,,\quad\text{at $P_\sigma\approx2.115\gamma_a$ when
    $\gamma_\sigma=\gamma_a$}\,.%\label{eq:SunMay22234754CEST2011}
\end{align*}
It is difficult to know where to place the threshold in the
one-emitter laser, other than the rather vague statement that it is
zero, which does not account well for the variety of situations that
can be observed. An unambiguous definition could be that point where
$g^{(2)}$ achieves its maximum, now that we have shown this is a
universal feature of lasing in strong coupling. In this case, there is
no ideal thresholdless laser and the lowest possible threshold is that
given by the condition that maximises strong-coupling,
$\gamma_\sigma=\gamma_a$, yielding a threshold at a pumping rate
slightly over twice this common decay
rate. %Eq.~(\ref{eq:SunMay22234754CEST2011}).

Beyond the two particular cases that we have just outlined, there lie
all the possible values of $\gamma$. From the maximum $g^{(2)}$
obtained, given that it is universal, one can also estimate the the
pumping rate and the imbalance of the decay rates, quantities
otherwise difficult to access directly. Interestingly, such a local
maximum of statistics when crossing the thresholds to stimulated
emission lasing have been observed in experimental realizations of a
few-emitters laser with a shape that resembles our
Fig.~\ref{fig:TueMay31115403CEST2011}~\cite{strauf06a,xie07a,ulrich07a,witzany11a},
but it was in all cases linked to an experimental limitation, whereas
it is in our case a manifestation of an intrinsic and universal
transition in the system. In the light of our findings, this
transition region acquires a new interest since it will allow
fundamental tests of the theory at the interface between quantum and
classical regimes, provide an unambiguous characterization of lasing
in strong coupling, quantify the extent of experimental limitations,
give a direct access to underlying parameters of the system and set
the lowest thresholds achievable in any device relying on strong
coupling.

\begin{acknowledgments}
  We thank A. Laucht for discussions. We acknowledge support from DFG
  via SFB-631, Nanosystems Initiative Munich, the Emmy Noether project
  HA 5593/1-1 funded by the German Research Foundation (DFG) and the
  Marie Curie IEF `SQOD'.
\end{acknowledgments}

\bibliography{Sci,lsc,books,arXiv}

\end{document}